# An infrared integrated optic astronomical beam combiner for stellar interferometry at 3-4 μm

Hsien-kai Hsiao,[1,*] K. A. Winick,[1] John D. Monnier,[2] and Jean-Philippe Berger[3]

[1]*Department of Electrical Engineering and Computer Science, University of Michigan, 1301 Beal Avenue, Ann Arbor, MI 48109, USA*
[2]*Department of Astronomy, University of Michigan, 500 Church Street, Ann Arbor, MI 48109, USA*
[3]*Laboratoire d'Astrophysique UMR CNRS/UJF 5571, Observatoire de Grenoble, BP. 53, F-38041 Grenoble Cedex 9, France*
[*]*hkhsiao@umich.edu*

**Abstract:** Integrated-optic, astronomical, two-beam and three-beam, interferometric combiners have been designed and fabricated for operation in the L band (3 μm – 4 μm) for the first time. The devices have been realized in titanium-indiffused, x-cut lithium niobate substrates, and on-chip electro-optic fringe scanning has been demonstrated. White light fringes were produced in the laboratory using the two-beam combiner integrated with an on-chip Y-splitter.

**References and links**

1. Max Born and Emil Wolf, *Principles of Optics*, 7[th] *(Expanded) Edition* (Cambridge University Press, 1999), Chap. 10.
2. A. A. Michelson, "Measurement of Jupiter's satellites by interference," Nature (London) **45**, 160 (1891).
3. Theo ten Brummelaar, Michelle Creech-Eakman, and John Monnier, "Probing stars with optical and near-IR interferometry," Physics Today, pp. 28-33 (June 2009).
4. Jonathan I. Lunine, Bruce Macintosh, and Stanton Peale, "The detection and characterization of exoplanets," Physics Today, pp. 46-51 (May 2009).
5. Jim Breckinridge and Chris Lindensmith, "The astronomical search for origins," Optics and Photonics News **16**, 24-29 (2005).
6. C.V.M. Fridlund, "The Darwin mission," Advances in Space Research **34**, 613-617 (2004).
7. NASA Exoplanet Community Report, Ed. P.R. Lawson, W. A. Traub and S. C. Unwin, JPL Publication 09-3 (2009). http://exep.jpl.nasa.gov/documents/ExoplanetCommunityReport.pdf
8. R. N. Bracewell, "Detecting nonsolar planets by spinning infrared interferometers," Nature **274**, 780-781 (1978).
9. P.R. Lawson, O.P. Lay, S.R. Martin, R.D. Peters, R.O. Gappinger, A. Ksendzov, D.P. Scharf, A.J. Booth, C.A. Beichman, E. Serabyn, K.J. Johnston, and W.C. Danchi, "Terrestrial planet finder interferometer 2007-2008 progress and plans," Proc. SPIE **7013**, 70132N (2008).
10. W.A. Traub, K.W. Jucks, and C. Noecker, "Biomarkers on extrasolar terrestrial planets: estimates of detectability," American Astronomical Society, 197th AAS Meeting, #49.03; Bulletin of the Ameri. Astro. Society **32**, 1485 (2000).
11. S. Brustlein, L. Del Rio, A. Tonello, L. Delage, F. Reynaud, H. Herrmann, and W. Sohler, "Laboratory demonstration of an infrared-to-visible up-conversion interferometer for spatial coherence analysis," Phys. Rev. Lett. **100**, 153903 (2008).
12. Kern P., Malbet F., Schanen-Duport I., and Benech P., "Integrated optics single-mode interferometric beam combiner for near infrared astronomy," in Proceedings of AstroFib '96, Integrated Optics for Astronomical Interferometry, pp. 195-204 (1996).
13. J.P. Berger, K. Rousselet-Perraut, P. Kern, F. Malbet, I. Schanen-Duport, F. Reynaud, P. Haguenauer, and P. Benech, "Integrated optics for astronomical interferometry. II. First laboratory white-light interferograms," Astron. Astrophys. Suppl. Ser. **139**, 173-177 (1999).
14. P. Haguenauer, J.P. Berger, K. Rousselet-Perraut, P. Kern, F. Malbet, I. Schanen-Duport, and P. Benech, "Integrated optics for astronomical interferometry. III. Optical validation of a planar optics two-telescope beam combiner," Appl. Opt. **39**, 2130-2139 (2000).
15. J.P. Berger, P. Haguenauer, P. Kern, K. Rousselet-Perraut, F. Malbet, I. Schanen, M. Severi, R. Millan-Gabet, and W. Traub, "Integrated optics for astronomical interferometry. IV. First measurements of stars," Astron. Astrophys. **376**, L31-L34 (2001).

16. E. Laurent, K. Rousselet-Perraut, P. Benech, J.P. Berger, S. Gluck, P. Haguenauer, P. Kern, F. Malbet, and I. Schanen-Duport, "Integrated optics for astronomical interferometry. V. Extension to the *K* band," Astron. Astrophys. **390**, 1171-1176 (2002).
17. J.B. LeBouquin, P. Labeye, F. Malbet, L. Jocou, F. Zabihian, K. Rousselet-Perraut, J.P. Berger, A. Delboulbé, P. Kern, A. Glindemann, and M. Schöeller, "Integrated optics for astronomical interferometry. VI. Coupling the light of the VLTI in *K* band," Astron. Astrophys. **450**, 1259-1264 (2006).
18. B. Mennesson, J.M. Mariotti, V. Coudé du Foresto, G. Perrin, S. Ridgway, W.A. Traub, N.P. Carleton, M.G. Lacasse, and G. Mazé, "Thermal infrared stellar interferometry using single-mode guided optics: first results with the TISIS experiment on IOTA," Astron. Astrophys. **346**, 181-189 (1999).
19. Guangyu Li, Tobias Eckhause, Kim A. Winick, John D. Monnier, and J.P. Berger, "Integrated optic beam combiners in lithium niobate for stellar interferometer," Proc. SPIE **6268**, 626834 (2006).
20. R.A. Becker, R.H. Rediker, and T.A. Lind, "Wide-bandwidth guided-wave electro-optic intensity modulator at $\lambda$ = 3.39 $\mu$m," Appl. Phys. Lett. **46**, 809-811 (1985).
21. John D. Monnier, "Asymmetric beam combination for optical interferometry," Publications of the Astronomical Society of the Pacific **113**, 639-645 (2001).
22. Rostislav Vatchev Roussev, "Optical-frequency mixers in periodically poled lithium niobate: materials, modeling and characterization," Ph.D. thesis, Stanford University (2006).
23. Kin Seng Chiang, "Construction of refractive-index profiles of planar dielectric waveguides from the distribution of effective indexes," Journal of Light. Tech. **LT-3**, 385-391 (1985).
24. E. Strake, G. P. Bava, and I. Montrosset, "Guided modes of Ti:LiNbO$_3$ channel waveguides: a novel quasi-analytical technique in comparison with the scalar finite-element method," Journal of Light. Tech. **6**, 1126-1135 (1988).
25. K. K. Wong, *Properties of Lithium Niobate*, (INSPEC, 2002), Chap. 8.
26. William J. Minford, Steven K. Korotky, and Rod C. Alferness, "Low-loss Ti:LiNbO$_3$ waveguide bends at $\lambda$ = 1.3 $\mu$m," IEEE Journal of Quant. Elect. **QE-18**, 1802-1806 (1982).
27. Yohei Sakamaki, Takashi Saida, Munehisa Tamura, Toshikazu Hashimoto, and Hiroshi Takahashi, "Low loss crosstalk waveguide crossings designed by wavefront matching method," IEEE Photon. Technol. Lett. **18**, 2005-2007 (2006).
28. R.C. Alferness, R.V. Schmidt, and E.H. Turner, "Characteristics of Ti-diffused lithium niobate optical directional couplers," Appl. Opt. **18**, 4012-4016 (1979).
29. K. Okamoto, *Fundamentals of Optical Waveguides*, (Academic Press, San Diego, 2000), Chap. 4.
30. K. Kishioka, "A design method to achieve wide wavelength-flattened responses in directional coupler-type optical power splitters," J. Lightwave Technol. **19**, 1705-1715 (2001).
31. Rober O. Gappinger, Rosemary T. Diaz, Alexander Ksendzov, Peter R. Lawson, Oliver P. Lay, Kurt M. Liewer, Frank M. Loya, Stefan R. Martin, Eugene Serabyn, and James K. Wallace, "Experimental evaluation of achromatic phase shifters for mid-infrared starlight suppression," Appl. Opt. **48**, 868-880 (2009).
32. Joss Bland-Hawthorn, Pierre Kern, "Astrophotonics: a new era for astronomical instruments," Optics Express **17**, 1880-1884 (2009).
33. N. Ho, M. C. Phillips, H. Qiao, P. J. Allen, K. Krishnaswami, B. J. Riley, T. L. Meyers and N. C. Anheier, Jr., "Single-mode low-loss chalcogenide glass waveguides for mid-infrared," Opt. Lett. **31**, 1860-1862 (2006).
34. L. Labadie, L. Abel-Tiberini, E. LeCoarer, C. Vigreuz-Bercovici, B. Arezki, M. Barillot, J.-E. Broquin, A. Delboulbé, P. Kern, V. Kirschner, P. Labeye, A. Pradel, C. Ruilier, and P. Saguet, "Recent progress in mid infrared integrated optics for nulling interferometry," Proc. SPIE **6268**, 62682E (2006).

## 1. Introduction

It is well known that high-angular resolution astronomical imaging can be done using interferometry. The Michelson stellar interferometer measures the complex amplitude correlation between two optical fields measured at spatially separated apertures. According to the van Cittert-Zernike theorem, this correlation is proportional to the Fourier transform of the object's intensity distribution at a spatial frequency equal to the projected separation of the apertures divided by the observing wavelength [1]. By changing the baseline of the apertures, the Fourier transform of the object's intensity distribution may be obtained at a large number of spatial frequencies, thus allowing an image of the object to be reconstructed. In 1890 on the twelve-inch aperture of a telescope at the Lick observatory, Michelson used stellar interferometry to measure the diameter of the four moons of Jupiter [2]. Recently, stellar interferometry has been applied to other areas of astrophysics such as high resolution imaging of binary stars, direct imaging of rapid rotating stars, the observation and modeling of circumstellar dust shells and the search for exoplanets [3].

There is a longstanding quest to find planets orbiting other stars (i.e., exoplanets), especially Earth-like planets in habitable zones where the surface temperature is able to support liquid water over a range of surface pressures. Since 1995, more than 350 planets beyond our solar system have been discovered indirectly by Doppler spectroscopy, astrometry, transit methods, and microlensing effects [4]. Most of the exoplanets discovered so far are giants similar to Jupiter and Saturn, which are unfavorable for supporting life. Space-based interferometry is a promising method to search and to characterize earth-like planets at high contrast and high angular resolution. NASA's Terrestial Planet Finder (TPF) program [5] and ESA's Darwin mission [6] are ambitious efforts to realize this goal. The main difficulty in imaging is separating the faint light of the planet from the bright emission of its host star. At mid-infrared wavelengths (7 μm -20 μm) a planet is over $10^6$ times fainter than its host star, with this ratio increasing to $10^{10}$ in the visible band. The TPF program spans several different mission concepts, including internal coronagraphs (TPF-C), external occulters (TPF-O) and nulling interferometers (TPF-I) [7]. In 1978, Ronald Bracewell proposed the use of a nulling interferometer to cancel the light coming directly from a bright star, thus making it possible to see relatively faint orbiting planets [8]. The TPF-I mission's goal is to build and deploy a mid-infrared, space-based, nulling interferometer based on Bracewell's basic idea. The space-based interferometer will be designed to find and measure the mid-infrared spectra of the atmospheres of Earth-like exoplanets around nearby stars. The TPF-I mission will search for evidence of key biomarkers, such as ozone, water, and carbon dioxide so that the possible presence of planetary life can be inferred [9,10].

High resolution interferometric imaging requires that more than two apertures be combined. One of the major advantages of integrated optic (IO) beam combiners, as opposed to purely optical fiber implementations, is the ability to combine multiple apertures on a single chip in a scalable manner. IO interferometers offer additional advantages over bulk implementations, including spatial filtering, enhanced stability, electrically-controlled, on-chip, phase modulation, and wavelength conversion [11]. IO beam combiners for astronomical imaging were first proposed by Kern et al. in 1996 [12]. Using silicate-based glass IO devices, laboratory and on-sky measurements of stellar interferograms were demonstrated by Berger et al. at astronomical H (1.5-1.8 μm) and K (2.0-2.4 μm) bands [13-17]. Star-to-planet brightness ratios make operation in the infrared beyond 3 μm attractive, but classical silicate-based glasses are not transparent in this spectral band. On-sky interferometric measurements have been performed in the L band using a two-beam fluoride glass fiber coupler, but this fiber-based technology is not easily scalable to multiple apertures [18].

Some preliminary measurement results for an integrated optic beam combiner, made by annealed proton-exchanged (APE) waveguides in lithium niobate ($LiNbO_3$), have been previously reported [19]. Electro-optic (EO) modulation in Ti-diffused $LiNbO_3$ at 3.39 μm has also been demonstrated [20]. In this paper, we describe the development of a prototype, single-mode, IO, astronomical beam combiner fabricated by titanium-indiffusion in x-cut $LiNbO_3$. The device operates in 3.2-3.8 μm region, which lies in the L band, and has on-chip, EO, controllable fringe scanning. Using a broadband thermal source in the laboratory, a white-light interferogram, along with on-chip EO fringe scanning, is demonstrated with an IO beam combiner operating in the L band for the first time. In section 2 the techniques used for device design and fabrication are presented. In section 3 laboratory measurements of a white-light interferogram in the L band are described. In section 4, our results are summarized and future work is discussed.

## 2. Design, fabrication, and characterization of the device

The overall size of the fabricated three-beam combiner chip, as illustrated in Fig. 1(a) is 60 mm (long) x 20 mm (wide) by 1 mm (thick). The device itself, excluding the electrical contact pads, occupies only 1.25 mm of the total chip width. A two-beam combiner, containing components similar to those contained in the three-beam device, is shown in Fig. 1(b). This later device was characterized in our laboratory. The beam combiners contain three optical functions: spatial filtering based on single-mode waveguides with sufficient length, interferometric beam combining based on 3 dB symmetric directional couplers, and electrically-controlled phase modulation based on thin film metal electrodes. For the three-beam combiner, the photometric signal levels can be determined using linear combinations of the interferometric outputs ($I_{12}+$, $I_{12}-$, $I_{13}+$, $I_{13}-$, $I_{23}+$, and $I_{23}-$) using a scheme similar to that reported in [21].

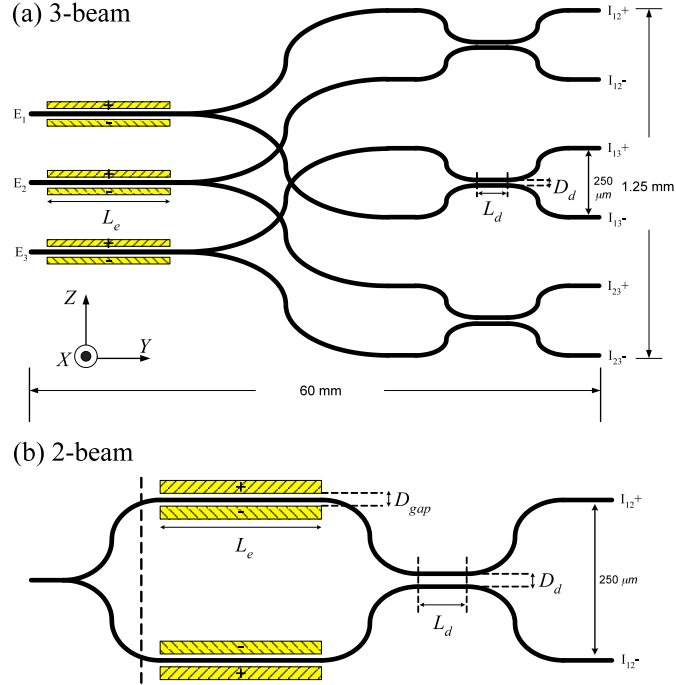

Fig. 1. Layout of (a) fabricated three-beam combiner (b) two-beam combiner for device characterization (figures not drawn to scale).

*2.1 Single-mode waveguide at a wavelength of 3.39 μm*

In order to obtain accurate visibility measurements the waveguides in the beam combiner must be single mode at the operating wavelengths. The beam combiner was fabricated in $LiNbO_3$ because this substrate has excellent transparency from visible wavelengths to approximately 4 μm, and it may be easily EO phase modulated [20]. There are two widely used methods for fabricating waveguides in $LiNbO_3$: annealed proton-exchanged (APE) and titanium in-diffusion (Ti:$LiNbO_3$). The later method produces low-index contrast waveguides, and thus the radius of curvature of the waveguide bends must be kept large in order to avoid significant bending losses. On the other hand, APE waveguides only support the mode polarized along the z-axis of the substrate, while both TE- and TM-modes can be supported by Ti:$LiNbO_3$ waveguides. Furthermore, APE waveguides show strong, broad OH absorption peaks around 3500 $cm^{-1}$ (~2.8 μm), which have the potential to significantly increase the propagation losses in 3-4 μm band [22]. Therefore the Ti-indiffusion method was chosen to fabricate the astronomical beam combiners reported here. Both planar and channel waveguides were obtained by diffusing a E-beam deposited, 1600Å thick Ti layer into congruent x-cut $LiNbO_3$ for 35 hours at 1050 ºC in a covered ceramic tray [20]. For the channel waveguides, the width, $W$, of the pre-diffused Ti strips was chosen to be 18 μm. In the y-z plane of the x-cut LiNbO$_3$ wafer, the channel waveguides were oriented parallel to the y-axis. A prism coupling technique was used to measure the effective indices of the TE planar waveguide modes at $\lambda$ = 0.633 μm. From the measured mode effective indices, a 1-D refractive index profile was constructed using the inverse WKB method devised by Chiang [23]. The resulting refractive index profile along with a corresponding Gaussian fit is shown in Fig. 2(b). From this data, the surface refractive index difference $\Delta n$ and 1/e diffusion depth $D_x$ were deduced to be 0.0136 and 8.0 μm, respectively. The effective indices were measured with a commercial (Metricon 2010) prism coupling instrument to a manufacturer's estimated accuracy of ±0.0005. The results are shown in Fig. 2(a) along with calculated effective indices assuming the Gaussian profile given in Fig. 2(b).

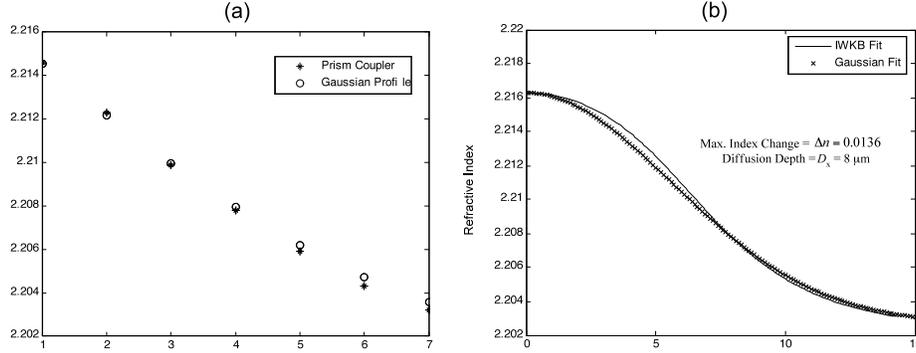

Fig. 2. (a) Comparison of the measured effective indices and calculated effective indices for a Gaussian profile (b) Extraordinary index profile of planar waveguide.

In order to estimate the channel waveguide modes and thus determine whether single-mode operation was possible, it was assumed that the 2-D refractive index profile $n(x,z)$ of the x-cut, Ti-indiffused channel waveguides could be approximated as follows [24]

$$n(x,z) = n_s + \Delta n \cdot \frac{1}{2}\left[erf\left(\frac{W/2+z}{D_z}\right) + erf\left(\frac{W/2-z}{D_z}\right)\right] \cdot \exp\left(-\frac{x^2}{D_x^2}\right) \quad (1)$$

In Eq. (1), $n_s$ is the substrate index and $W$ is the pre-diffusion width of the E-beam deposited Ti strips. Furthermore for simplicity, the lateral and longitudinal diffusion depths $D_z$ and $D_x$, respectively, were assumed to be equal. Using Eq. (1) and $W$ = 18 μm, the channel waveguide modes at $\lambda$ = 3.39 μm were determined by the beam propagation method. The $\Delta n$ and $D_x$ values used in this calculation were 0.0136 and 8.0 μm, respectively, as given by our planar measurements (at $\lambda$ =0.63 μm) described above. The substrate refractive index $n_s$ was taken to be 2.0823 at $\lambda$ = 3.39 μm for the TE mode [25], though the results are relatively independent of the exact value used. The beam propagation results indicated that the channel waveguides should only support a single TE and a single TM mode at $\lambda$ = 3.39 μm. Furthermore, the TM mode was very close to cut-off. The assumed refractive index profile given by Eq. (1) and the corresponding calculated TE mode profile are shown in the Figs. 3(a) and 3(b), respectively. In these figures, the air-LiNbO$_3$ boundary is located at x = 0 μm. Since we do not know the explicit wavelength dependence of $\Delta n$ much beyond 1.5 μm, our prediction that the waveguides will not be multi-moded at 3.39 μm relies on the expectation that $\Delta n$ does not increase with wavelength. Such an assumption is consistent with data presented for the visible and near IR bands out to approximately 1.5 μm [24].

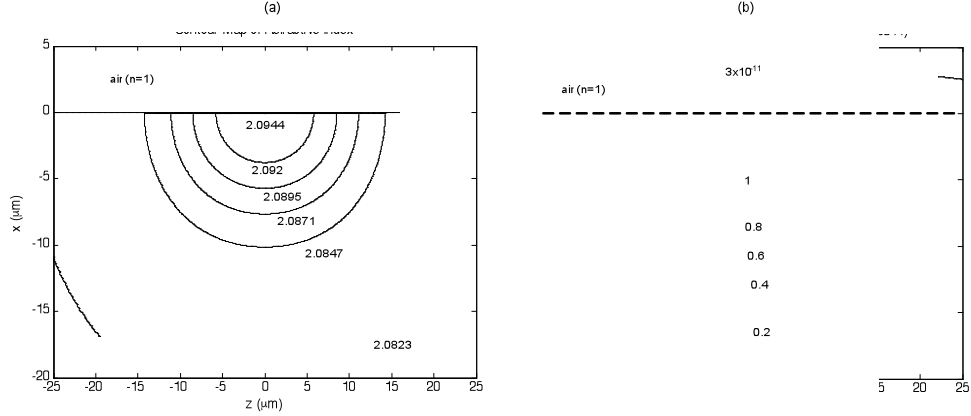

Fig. 3. (a) Contour map of Refractive index (b) Computed TE-mode profile.

For the fundamental TE-polarized mode at $\lambda = 3.39$ μm, the calculated effective index, $n_{eff}$(2.0844), lies 0.0021 above the substrate refractive index. A similar calculation shows that TM-polarized mode is near cut-off, and thus is expected to experience very high bending losses. The EO modulation results shown in Fig. 9 indicate that the waveguide is single-moded in the L band, since the output response is quasi-periodic. Different spatial modes are orthogonal and thus do not beat against one another. Consequently if there were two (or more) discrete modes, each one will separately produce a quasi-periodic intensity variation as the optical path length in one arm is varied. The period of this variation (corresponding to each spatial mode) will depend on the wavelength and the effective index of the mode. Since the two modes have different effective indices, two (or more) intensity variations with different periods will be superimposed. Such a superposition does not yield a quasi-periodic variation.

*2.2 Characterization of bending losses*

The relatively small increase of refractive index obtained by Ti-indiffusion leads to weak confinement of optical modes, and thus greatly increases the losses in the bending regions. Therefore, waveguide bends of large radius of curvature must be used to minimize bending losses and keep the device size reasonable.

In order to characterize the bending losses, a set of S-shaped waveguide bends consisting of two back-to-back semi-circular arcs of constant radius of curvature $R$ were fabricated as shown in Fig. 4. The losses, including pure bending losses and those arising from abrupt curvature changes due to waveguide-to-waveguide transitions, were measured by comparing the throughput of these S-shaped bends to straight waveguide sections. The results are shown in Fig. 5, along with an exponential fit, for values of $R$ between 6 and 22 cm. For circular arcs of radius $R$, the pure bending loss coefficient can be modeled as

$$\alpha(R) = C_1 \exp(-C_2 R) \quad \text{(dB/cm)} \qquad (2)$$

where parameters $C_1$ and $C_2$ are independent of $R$, but are functions of the waveguide parameters [26].

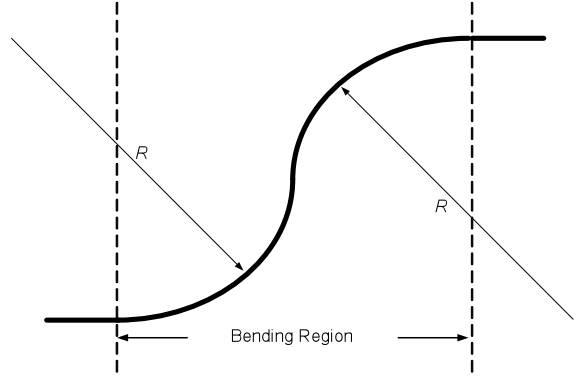

Fig. 4. Two semi-circular arcs with radius of curvature *R*.

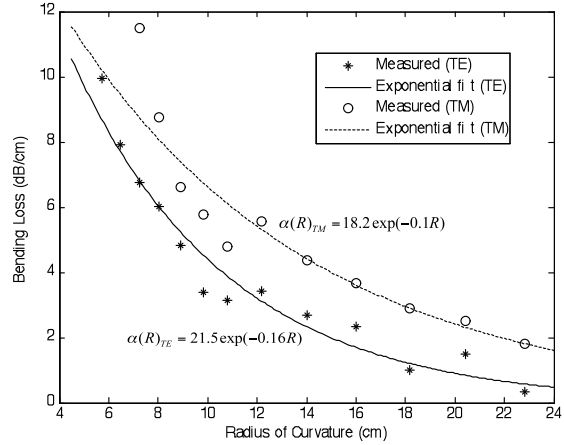

Fig. 5. Bending loss measurements and exponential fit.

Although not excellent, the quality of the exponential fit indicates that the losses are dominated by pure bending as opposed to transition losses. According to the fit shown in Fig. 5, the values of $C_1$ and $C_2$ at $\lambda = 3.39$ μm are 21.5 dB/cm and 0.16 cm$^{-1}$ for TE-mode and 18.2 dB/cm and 0.1 cm$^{-1}$ for TM-mode, respectively. Thus, the TM-mode experiences much higher losses. In order to minimize the TE-mode losses, the radius of curvature of the S-bends for the couplers was chosen to be as large as possible, i.e. 22 cm, while still allowing the 3-beam combiner to be fabricated using a 3 inch diameter LiNbO$_3$ wafer. The corresponding bending losses were 0.64 dB/cm and 2 dB/cm for the TE-mode and TM-mode, respectively.

By launching optical power, $P_1$, into a single input port of the two-beam combiner (without Y-splitter), the insertion loss of the device (i.e., $-10\log_{10}(P_{12+} + P_{12-})/P_1$) for the TE-mode was found to be 4.7 dB larger than that of a straight waveguide of the same length. We separately measured the excess loss of the two-beam integrated Y-splitter together with its associated waveguide bend and found it to be approximately 2 dB. Generally, Ti-indiffused LiNbO$_3$ waveguides exhibit low propagation loss (< 0.1 dB/cm). Thus the large losses reported here come primarily from the bending losses that are due to the weak confinement of the guided mode at $\lambda = 3.39$ μm. The use of larger LiNbO$_3$ wafers and/or thicker Ti-indiffused strips would allow us to increase the radii of curvature of the S-bends and/or to increase the mode confinement, greatly reducing insertion losses.

A waveguide intersected by another waveguide will perturb the electric field distribution of the modes, producing both excess loss and cross-talk [27]. Excess losses and cross-talk generally becomes more severe as the angle of crossing decreases. On the other hand, decreasing the crossing angle enables one to use bends with a larger radius of curvature and/or to build smaller devices. Thus some compromise between performance and size must be reached. For the 3-beam combiner shown in Fig. 1(a), the crossing angles were approximately 4°. A series of X-shaped waveguide crossings with different crossing angles were fabricated in order to estimate the cross-talk that the 3-beam combiner would experience. The cross-talk (i.e., output power from cross port/sum of output powers from the two output ports) for these X-shaped waveguide crossings was measured to be below 20 dB. The excess loss was not measured directly, but it is expected to be relatively small given the small value of the cross-talk.

*2.3 Directional coupler design*

For phase-matched operation of the directional coupler, the power transfer efficiency $\eta$ is given by the expression [28]

$$\eta = \sin^2(\kappa L_d + \varphi) \qquad (3)$$

where $\kappa$ is the coupling coefficient and $L_d$ is the interaction length of the straight section. The phase term $\varphi$ corresponds to the coupling in the S-bend regions where $\kappa$ is not a constant. A set of directional couplers with center-to-center spacing $D_d$ = 26 µm, S-bend radius of curvature $R$ = 6 cm, and coupling lengths $L_d$ ranging from 3 to 9 mm were fabricated. The power transfer efficiencies of these couplers were measured, and the results are plotted in Fig. 6 along with a fit to Eq. (3). From this fit, the value of $\kappa$ was found to be 0.48 mm$^{-1}$ for both the TE and TM fundamental modes.

For the final design with $R$ = 22 cm, power splitting ratios were measured for three different values of $L_d$ (3.34, 3.44, and 3.54 mm). By fitting these three data points to Eq. (3) with same $\kappa$ experimentally determined in Fig. 6, the $\varphi$ value corresponding to $R$ = 22 cm was found to be 0.95. Thus it follows from Eq. (3) that an interaction length of approximately $L_d$ = 2900 µm will correspond to the required 3 dB coupler. Using the HeNe laser operating at 3.39 µm, the measured power splitting ratio $\eta$ for TE-mode (TM-mode) of the directional coupler with $L_d$ = 2900 µm and $R$ = 22 cm was measured to be 48% (54%), corresponding to a fringe visibility of 99.9% (99.7%).

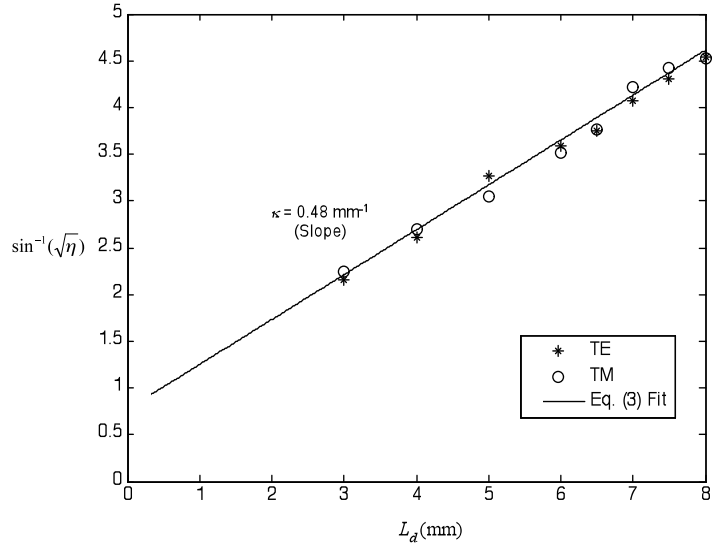

Fig. 6. Experimentally determined coupling characteristics.

*2.3.1 Wavelength-dependent response*

Long baseline, stellar interferometry measures fringe visibilities and phase closure. Deep nulling, as well as high contrast broadband fringe formation, requires that the beam combiner and fringe scanner operate achromatically. A lossless directional coupler is a four-port device. For monochromatic operation at wavelength λ, the squared-magnitudes of the complex-valued electric field amplitudes, $A_{out}$ and $B_{out}$, measured at the two output ports can be written as [29]

$$|A_{out}|^2 = |A_{in}|^2 [1-C(\lambda)] + |B_{in}|^2 C(\lambda) - 2|A_{in}| \cdot |B_{in}| \sqrt{C(\lambda)[1-C(\lambda)]} \cdot \sin(\angle A_{in} - \angle B_{in} + \phi_{int}(\lambda) + \theta_{ext}(\lambda))$$
$$|B_{out}|^2 = |A_{in}|^2 [1-C(\lambda)] + |B_{in}|^2 C(\lambda) + 2|A_{in}| \cdot |B_{in}| \sqrt{C(\lambda)[1-C(\lambda)]} \cdot \sin(\angle A_{in} - \angle B_{in} + \phi_{int}(\lambda) + \theta_{ext}(\lambda)) \quad (4)$$

where $A_{in}$ and $B_{in}$ are the complex-valued electric fields with amplitudes $|A_{in}|$, $|B_{in}|$ and phases $\angle A_{in}$, $\angle B_{in}$ measured at the two input ports, $C(\lambda)$ is the directional coupler's wavelength-dependent power cross-over ratio (i.e., $|B_{out}|^2/|A_{in}|^2$ when $B_{in}=0$), $\phi_{int}(\lambda)$ is the wavelength-dependent phase introduced by the directional coupler, and $\theta_{ext}(\lambda)$ is an externally-induced phase shift between the two input beams. For the EO scanning reported in this paper, the last of these quantities is given by

$$\theta_{ext}(\lambda) = \frac{2\pi}{\lambda} \Delta n_{eo}(\lambda) L_e \quad (5)$$

where $\Delta n_{eo}(\lambda)$ is the wavelength-dependent EO-induced refractive index change and $L_e$ is the physical path length of the EO waveguide section. For broadband sources, Eq. (4) must be integrated across the source spectrum to derive the final fringe intensity. The directional coupler will be achromatic if and only if $C$ and $\phi_{int}$ are not wavelength-dependent. It is also clear from Eq. (5) that the path-length difference introduced by EO scanning is wavelength-dependent.

Symmetrical directional couplers are not achromatic devices. Several techniques, including the use of asymmetrical couplers, have been proposed and demonstrated to realize "wavelength-flattened" operation [30]. It is interesting to note, however, that most proposals to realize wavelength-flattened device operation consider only the directional coupler's power transfer characteristic (i.e., $C(\lambda)$) and ignore the wavelength-dependent phase shift $\phi_{int}(\lambda)$. The phase shift determines the fringe position, and obtaining a wavelength-independent fringe position may be as important as realizing a wavelength-flattened intensity response. This is especially true for nulling applications. It is easily demonstrated that the phase response of a symmetrical directional coupler is wavelength-independent [29], while this is not the case for previously proposed wavelength-flattened, asymmetrical, directional couplers. The development of IO directional couplers and fringe scanners with enhanced achromaticity warrants further investigation.

*2.4 Phase modulator*

The push-pull configuration of electrodes was fabricated on the beam combiners using traditional photolithographic processing. 60 μm wide, 500 nm thick gold strips were evaporated next to the waveguides with an edge-to-edge electrode spacing $D_{gap}$ of 24 μm as shown in Fig. 1(b). The electrode lengths $L_e$ on both the two-beam and three-beam combiners were chosen to be 1 cm. The direction of the applied electric field lies primarily along the optical axis (z-axis) of $LiNbO_3$, and thus the largest EO coefficient $r_{33}$ was utilized for the TE-mode. The EO-induced phase shift $\Delta\varphi$ is given by Eq. (6)

$$\Delta \varphi = 2\Gamma_{eff} \cdot \pi \cdot n^3 \cdot r \frac{V}{D_{gap}} \frac{L_e}{\lambda} \qquad (6)$$

where $n$ is the refractive index of the LiNbO$_3$ substrate, $r$ is the EO coefficient ($r_{33}$ for the TE-mode and $r_{13}$ for the TM-mode) and $\Gamma_{eff}$ is factor between 0 and 1 that accounts for the spatial overlap between the applied electric field and the mode. A HeNe laser operating at 3.39 μm was launched into the Y-splitter of the two-beam combiner shown in Fig. 1(b), while a triangular-shaped voltage waveform was applied to the electrodes and the power coming from one output arm of the directional coupler was measured. The results are shown in Fig. 7 and correspond to a $V_\pi$ for the on-chip phase modulator of approximately 29 $V$/cm. This value is consistent with earlier reports [20], and corresponds to a $\Gamma_{eff}$ of approximately 0.5. Similarly, the measured $V_\pi$ (not shown) and the corresponding $\Gamma_{eff}$ for the TM-mode were found to be 165 V/cm and 0.25, respectively. It is worth noting that $\Gamma_{eff}$ is wavelength dependent. However, we are unable to characterize it due to lack of tunable IR source within the L band.

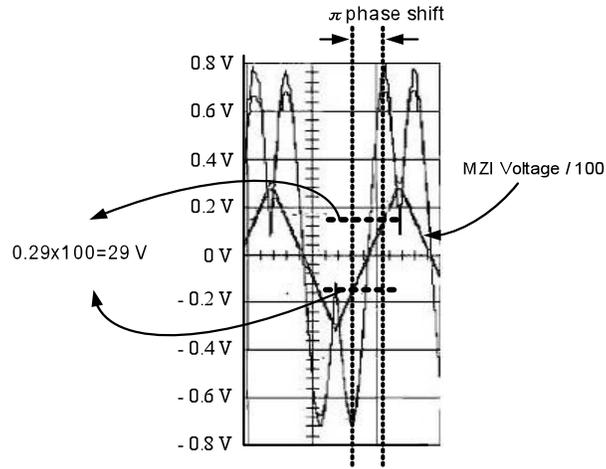

Fig. 7. The oscilloscope traces of applied MZI voltage and output signal power.

## 3. Laboratory white-light fringe measurement

The experimental set-up used to generate white-light fringes for the two-beam combiner in the laboratory is shown in Fig. 8. HeNe lasers at 633 nm and 3.39 μm were used to facilitate alignment so that maximum power from the white light source could be coupled into the on-chip Y-splitter. Using UV fused silica aspheric lenses with focal length 15 mm, an 8 W thermal source (Boston Electronics IR-12) operating at a blackbody temperature of 1170 K was coupled into the on-chip Y-splitter. The output of the directional coupler was imagined onto a liquid-nitrogen-cooled IR InSb detector after passing through a linear polarizer (TE) and a bandpass filter centered at 3.5 μm with a nearly 500 nm wide rectangular-shaped pass band. Due to the small amount of power that could be coupled into the waveguide from the broadband IR source, the signal at the InSb detector was buried in thermal noise and background light, and thus a lock-in amplifier was used to make the measurement. A DC voltage, which could be scanned from -175 V to +175 V, was applied to the electrodes. A small 1 KHz sinusoidal dither was added to this DC voltage, and a lock-in amplifier was used to synchronously measure the detector output. The value of the DC bias determined the position along the fringe.

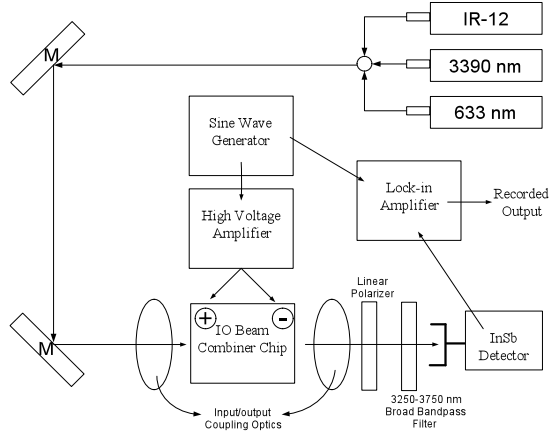

Fig. 8. Experimental set-up for measurement of white-light interferograms.

The measured interferogram (TE-mode) is shown in Fig. 9. Also shown in Fig. 9 is the theoretically predicted fringe given by Eq. (7) below

$$I_o = \frac{1}{\Delta\lambda}\int_{\lambda_0-\frac{\Delta\lambda}{2}}^{\lambda_0+\frac{\Delta\lambda}{2}} \frac{1}{2}\cos(\frac{2\pi\Delta nL_e}{\lambda})d\lambda \cong \frac{1}{2}\cos\Delta\varphi_0 \cdot \frac{\sin(\Delta\varphi_0 \cdot \frac{\Delta\lambda}{2\lambda_0})}{\Delta\varphi_0 \cdot \frac{\Delta\lambda}{2\lambda_0}} \tag{7}$$

where $\lambda_0 = 3.5$ μm, $\Delta\lambda = 500$ nm, and $\Delta\varphi_0 = 2\pi\Delta nL_e/\lambda_0 = (\pi V)/V_\pi(\lambda_0)$ is the EO induced phase shift at $\lambda_0$. In deriving Eq. (7) the wavelength-dependent behavior of the directional coupler has been ignored, i.e., $C(\lambda)$ in Eq. (4) has been assumed to be equal to 0.5 independent of wavelength. In Fig. 9 the $V_\pi$ for TE-mode was chosen to be 26.2 V, which is close to the value previously reported in section 2.4, and gives the best fit to the experimental data. Based on the power splitting ratio reported in section 2.3, the fringe visibility at 3.39 μm is approximately 99.9%. The lock-in measurement technique did not permit us to measure the *absolute* value of the fringe visibility when using the broadband thermal source. At the present time, we cannot fully explain the deviation of measured data from the theoretical fit at the larger electrode voltages. The wavelength dependence of the coupler as well as the wavelength dispersion of the EO-induced refractive index change would not result in the asymmetry of the fringes shown in Fig. 9.

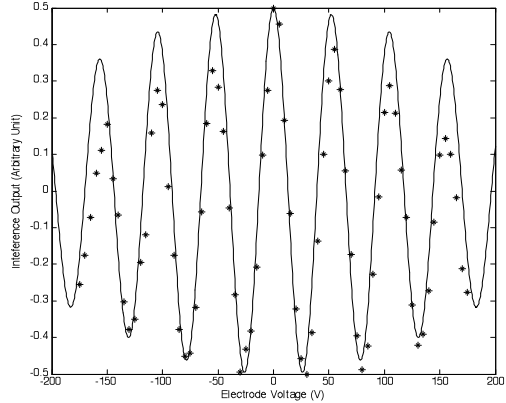

Fig. 9. White-light fringe and theoretical fringe with 500 nm bandwidth.

## 4. Conclusions and future work

Integrated-optic, astronomical, two-beam and three-beam, interferometric combiners have been designed and fabricated for operation in the L band, and the broadband on-chip electro-optic fringe scanning in titanium-indiffused, x-cut lithium niobate waveguides has been demonstrated for the first time. White-light fringes (TE-polarization) were produced in the laboratory using the two-beam combiner integrated with an on-chip Y-splitter. Using an applied electrode voltage of 350 V, it was possible to scan through approximately six fringes.

The devices exhibited relatively high waveguide losses in the bending regions due to the low-index contrast of the waveguides. Loss reduction is a key aspect of future development. Further optimization of the device layout, the use of larger wafers and/or thicker indiffused titanium strips is expected to reduce the losses to reasonable values. It should also be possible, using on-chip, polarization splitters to design devices that utilize both the TE and TM polarized light, thus capturing 3 dB more signal power. As discussed in section 2.3.1, the response of directional couplers and EO-induced fringe scanners is generally not achromatic. Techniques need to be developed to suppress the wavelength-dependent operation of these devices, especially for applications that require deep nulls [31].

Efforts are on-going in our laboratory to characterize the three-beam combiner using a thermal source and an astronomical InSb focal plane array. This three-beam combiner (shown in Fig. 1) is equivalent to the most complex IO combiner that has been used for on-sky measurements. Its on-sky measurement potential, however, surpasses previous IO combiners because it operates in the L band and has on-chip fringe scanning capability. The recent and widespread introduction of photonic devices for astronomical instrumentation has led to a new field and the associated neologism *astrophotonics* [32]. The field of mid-IR integrated optics is still in its infancy. There is every reason to believe, however, that the IO results presented in this paper can be extended beyond 4 μm to the mid-IR (7 μm - 20 μm). This effort will require substrate materials that are transparent at mid-IR, such as the chalcogenides [33,34]. Future work will explore these longer wavelength materials, associated waveguide fabrication methods, and new broadband, achromatic, IO techniques for next generation stellar interferometry and nulling.

## Acknowledgments

This work was performed under grant NNG04GC00G from the National Aeronautics and Space Administration. Special thanks are due to Dr. Guangyu Li of Electro Scientific Industries for helpful discussions.